\documentstyle[12pt,epsfig]{article}

\newcommand{\be}{\begin{equation}}
\newcommand{\ee}{\end{equation}}
\newcommand{\g}{{\bf g}}

\newcommand{\bbf}{\bf}
\newcommand{\ssl}{\sl}

\newcommand{\bea}{\begin{eqnarray}}
\newcommand{\eea}{\end{eqnarray}}

\begin{document}

\begin{titlepage}

\

\vspace{3 cm}

\begin{center}

{\Huge Gravitating Yang-Mills vortices in 4+1 spacetime dimensions}

\vspace{2 cm}

{\bf Mikhail S. Volkov}%
\footnote{Former address:
Institute for Theoretical Physics, University of Jena,
Max-Wien Platz 1, D-07743, Jena,
Germany}

\vspace{2 cm}

{\sl Laboratoire de Math\'ematiques
et Physique Th\'eorique,\\
Universit\'e de Tours,
Parc de Grandmont, \\
37200 Tours, 
FRANCE
}

\end{center}

\vspace{1 cm}

The coupling to
gravity in D=5 spacetime 
dimensions is considered for the particle-like and
vortex-type solutions  obtained by uplifting the
D=4 Yang-Mills instantons and D=3 Yang-Mills-Higgs monopoles. 
It turns out that the particles become
completely destroyed by gravity, while
the vortices admit a rich spectrum of
gravitating generalizations. 
Such vortex defects may be interesting in view of the AdS/CFT
correspondence or in the context of the brane world scenario. 

\end{titlepage}

\noindent
{\bf 1. Introduction.--}
The recent interest in
the AdS/CFT correspondence (see \cite{Aharony99} for a review), 
and in the brane world scenario \cite{Randall99}
has attracted attention to gravity theories in D=5 
spacetime dimensions.
In this connection solutions of gauged supergravities have been
actively studied. As   
such theories generically contain the non-Abelian gauge fields,
considering solutions for such fields coupled to gravity seems to be
important. At the same time, 
most studies so far have been restricted exclusively to the 
Abelian sector. Practically all what is known about gravitating
non-Abelian solutions in D=5 are the BPS configurations described 
in \cite{Gibbons94a}.
In D=4 on the other hand, 
gravitating Yang-Mills (YM) fields have been analyzed in some detail
(see \cite{Volkov98} for a review), partially with the 
numerical methods.  

The aim of this letter is to 
study more systematically gravity-coupled YM fields in D=5 by 
applying the methods previously used  in D=4. 
Instead of specializing to 
any particular supergravity model, we shall consider   
the pure Einstein-YM (EYM) theory. Although this theory 
is probably non-supersymmetric in itself, 
it  enters all gauged supergravities as the basic building block, and
hence one can expect some features of its solutions to be generic. 
In principle, one can study this theory
also in higher dimensions, but we shall work only in D=5. 

As is well known, the pure YM theory in D=5 Minkowski space 
admits topologically stable, particle-like  and
vortex-type solutions obtained by uplifting the D=4 YM instantons and
D=3 YM-Higgs monopoles. It is then natural to wonder
what happens to these objects when they are coupled to gravity. 
This issue seems to be interesting in itself, and it has not been 
addressed before. 
Below we shall study this problem and find  
that the particles become completely destroyed 
by gravity, as a result of their scaling behavior, 
while the vortices admit very non-trivial gravitating generalizations.
These gravitating vortices comprise an infinite family 
including the fundamental solution and its excitations. 
All solutions exist for a finite range of values of the 
gravitational coupling constant,  
and in the strong gravity limit they become 
gravitationally closed. 

It seems plausible that such solutions could be further 
generalized to include scalars and a cosmological term. In this case they
would describe stable vortex excitations over the AdS$_5$ background. 
Such topological defects might be interesting 
in view of the AdS/CFT correspondence 
as providing the dual gravity 
description for some processes on the gauge 
theory side, or possibly in the context of the 
brane world scenario. 

\noindent
{\bf 2. The model.--}
The pure
Einstein-YM theory for the gauge group
SU(2) is defined by the action
\be                   \label{1}
S=\int\left(-\frac{1}{16\pi G}\,R
-\frac{1}{4g^2}\,F^a_{MN}F^{aMN}\right)\sqrt{\g}\,d^5 x\, .
\ee
Here $F^a_{MN}=\partial_M A^a_N-\partial_N A^a_M
+\varepsilon_{abc}A^b_M A^c_N$ $(a=1,2,3)$,
and
$[G^{1/3}]=[g^2]=[{\rm length}]$.
Let us split the coordinates as
$x^M=(x^0,x^\mu)$, where $x^\mu=(x^i,x^4)$.
In the flat spacetime limit, $G\to 0$, the YM theory
is not conformally invariant, the length scale being $g^2$,
and this allows for soliton solutions. 
Specifically, for static, purely magnetic fields with
$A^a_M=(0,A^a_\mu(x^\nu))$ the energy
$E=\frac{1}{4g^2}\int (F^a_{\mu\nu})^2 d^4 x$
coincides with the action of the D=4 Euclidean YM theory.
It follows then that there are regular,
topologically stable solutions in D=5
with the energy $E=8\pi^2|n|/g^2$, where the topological
winding number $n\in\pi_3$(SU(2)).
These solutions describe neutral, particle-like objects 
which we shall call
``YM instanton particles''. 
If $\partial/\partial x^4$ is a symmetry,
one can choose $A^a_M=(0,A^a_i(x^k),H^a(x^k))$,
and the energy per unit $x^4$,
$E=\frac{1}{2g^2}\int(
(\partial_iH^a+\varepsilon_{abc}A^b_i H^c)^2+\frac12 (F^a_{ik})^2)d^3 x$,
coincides with the energy  of the D=3 YM-Higgs fields.
Since $H^aH^a$ is asymptotically constant \cite{Jaffe80},
there are topological vortex-type solutions with the  energy
per unit length $E=4\pi|n|/g^2$, where $n\in\pi_2(S^2)$.
These we shall call  ``YM vortices''. 
Let us now set $G\neq 0$.

\noindent
{\bf 3. Gravitating YM particles.--}
We parameterize the SO(4)-invariant line element 
in the one-particle sector as
\be                  \label{2}
ds^2=\sigma^2 Ndt^2
-\frac{d r^2}{N}-
r^2\,d\Omega_3^2\, ,
\ee
with $N\equiv 1-\kappa m/r^2$, 
the length scale being $g^2$.
The 
gauge field is given by $A^a=(1+w)\,\theta^a$.
Here $w$, $\sigma$, $m$ are functions of $r$,
$d\Omega_3^2=\theta^a\theta^a$ is the line element of the unit
sphere $S^3$, and $\theta^a$ are
the invariant forms on $S^3$, such that
$d\theta^a+\varepsilon_{abc}\,\theta^b\wedge\theta^c=0$.
The gravitational coupling constant is $\kappa=8\pi G/g^6$.
The  ADM mass, $M=m(\infty)$, determines the dimensionful energy
$E_{\rm ADM}=(3\pi^2/g^2) M$.
The independent field equations read
\bea          
&&r^2 Nw''+r\,w'
+\kappa\,(m-(w^2-1)^2)\frac{w'}{r}=2\,(w^2-1)w\, ,\nonumber \\
&&rm'=r^2Nw'^2+(w^2-1)^2\, ,   \label{15}
\eea
while the equation for $\sigma$ decouples
and can be integrated giving 
$\sigma(r)=\exp\left(-\kappa\int_r^\infty\frac{w'^2}{r}\,dr\right)$.
It follows that $r(m\sigma)'=(r^2w'^2+(w^2-1)^2)\sigma$, which
can be integrated with the
boundary conditions  $m(0)=0$ 
to give
\be                 \label{18}
M[w,\kappa]=\int_0^\infty\frac{dr}{r}\,(r^2w'^2+(w^2-1)^2)\,
\exp\left(-\kappa\int_r^\infty\frac{w'^2}{r}\,dr\right).
\ee
\begin{figure}
\epsfxsize=9cm
\epsfysize=7cm
\centerline{\epsffile{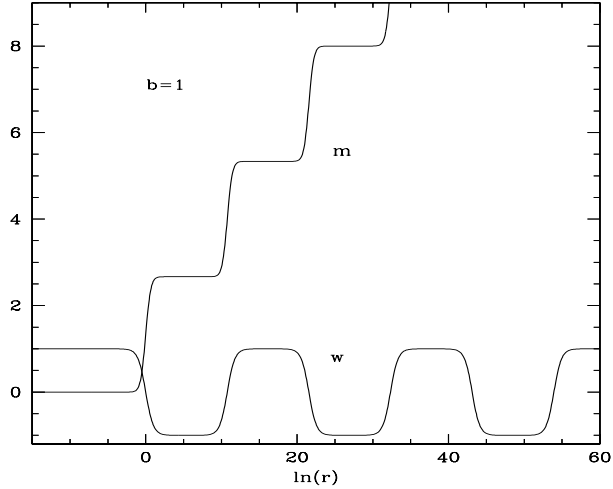}}
\caption{Typical solution to Eqs.(\ref{15}) (with $\kappa=10^{-8}$)}
\label{figYM5}
\end{figure}
This is the ADM mass for configurations subject
to the $(00)$ and $(rr)$ Einstein equations,
and for such fields it is proportional to the action density. 
It follows that on-shell fields are stationary points of $M[w,\kappa]$. 
For $\kappa=0$ the integrand in (\ref{18})
can be represented as a sum of a total derivative and a
perfect square. The latter vanishes if the self-duality equation holds,
$rw'=\pm(w^2-1)$,
whose solutions are 
$w=\mp\frac{b\,r^2-1}{b\,r^2+1}$, where $b$ is a
scale parameter. These describe the regular ``instanton particles''
with the energy 
$E_{\rm ADM}=8\pi^2/g^2$.

It turns out that for $\kappa\neq 0$ Eqs.(\ref{15}) do not 
admit globally regular solutions with finite ADM mass.
Indeed, for such solutions $\sigma(r)\neq 0$, such that the exponent
in (\ref{18}) never vanishes,   
and therefore (\ref{18}) can only be finite if
$w(r)\to\pm 1$ for $r\to 0,\infty$. It follows then that 
$M[w(\lambda r),\kappa]=M[w(r),\lambda^2\kappa]$,
where $\lambda$ is a constant scaling parameter.  
Since $M$ should be stationary with respect to rescalings, 
one should have
$0=\frac{d}{d\lambda}M[w(r),\lambda^2\kappa]|_{\lambda =1}$, 
but the latter is manifestly negative for $\kappa\neq 0$.

The YM particles therefore do not generalize to curved space --
as a consequence of their scaling behavior. Specifically, since the
size of the particles can be arbitrary,  
there is the corresponding scaling zero mode of the energy 
functional. This does not 
allow one to apply the inverse function theorem 
\cite{Kastor92} to construct the gravitating generalizations
even for arbitrarily small $\kappa$. 
Qualitatively, the YM particles resemble dust, since their 
rescaling changes the gravitational binding
(the exponential factor in (\ref{18})), without
affecting the energy of the gauge field
(the rest of the integrand in (\ref{18})).
As a result,  the  attraction and repulsion are not balanced
and the equilibrium is impossible -- the gravitating system wants
to shrink but there is no pressure to stop the contraction. 
Notice that these arguments apply only in D=5. 
In D=4 the YM instantons can be coupled to
gravity without problems -- 
since their energy-momentum tensor then vanishes,
they fulfill the coupled system of the Euclidean 
EYM equations for
any value of $G$. 

One can numerically integrate  Eqs.(\ref{15}) 
with the regular boundary conditions at the origin,
$w=1-2br^2+O(r^4)$, $m=O(r^3)$, in order to see what actually happens
to the static YM particle solution if $\kappa\neq0$. 
It turns out that as $r$ increases, 
$w$ first follows very closely (for small $\kappa$)
the flat space configuration, but then it fails to reach the value $-1$ 
as $r\to\infty$ and starts replicating the same pattern infinitely many 
times. As a result there emerges the quasi-periodic 
structure shown 
in Fig.\ref{figYM5}.  
This can be explained:
integrating  the Yang-Mills equation in (\ref{15})
gives
\be               \label{11}
w^{\prime 2}-(w^2-1)^2=
-\int_{-\infty}^\tau(\ln(\sigma^2 N))'w'^2 d\tau\, ,
\ee
with $^\prime \equiv\frac{d}{d\tau}\equiv r\sqrt{N}\frac{d}{dr}$.
This describes a particle moving with friction
in the inverted double-well potential $U=-(w^2-1)^2$. 
At $\tau=-\infty$ the particle
starts at the local maximum of the potential
at $w=1$, but then it looses energy 
due to the dissipation and
cannot reach the second maximum at $w=-1$. 
Hence it bounces back.
For large $\tau$ 
the dissipative term tends to zero, 
and the particle ends up oscillating in the
potential well with a constant energy. After each oscillation the mass
function $m$ increases in a step-like fashion, and for large $r$
one has $m\sim \tau\sim\ln r$. As a result, 
there emerges an infinite sequence of static spherical shells of the 
YM energy in the D=5 spacetime.

\noindent
{\bf 4. Gravitating YM vortices.--}
We parameterize the SO(3)-invariant metric
in the one-vortex sector as
\be                  \label{2:1}
ds^2={\rm e}^{2\nu}dt^2
-{\rm e}^{2\lambda}d r^2-
{\rm e}^{2\mu}\,d\Omega_2^2-{\rm e}^{2\zeta}\,(dx^4)^2,
\ee
with $d\Omega_2^2=d\vartheta^2+\sin^2\vartheta d\varphi^2$.
The SO(3)-invariant gauge field  
is given by
\be              \label{gauge}
A^a=
(w-1)\,\varepsilon_{abc}\,n^b dn^c+Hn^a\,dx^4\, ,
\ee
where $n^a=(\sin\vartheta\cos\varphi,\sin\vartheta\sin\varphi,\cos\vartheta)$.
Here $\nu$, $\lambda$, $\mu$, $\zeta$, $w$, and $H$ are functions of $r$. 
With $s={\rm e}^{\nu+\zeta+2\mu-\lambda}$
the field equations read
\bea             
&&s\,(s\,{\rm e}^{-2\mu}w')'-{\rm e}^{2\nu+2\zeta}(w^2-1)w=
{\rm e}^{2\nu+2\mu}H^2w \, ,       \label{2:7}  \\
&&s\,(s\,{\rm e}^{-2\zeta}H')'=
2{\rm e}^{2\nu+2\mu}w^2H \, ,      \label{2:7a} \\
&&s\,(s\,\nu')'=\frac{\kappa}{3}\,(A+2B+2C+D)\, ,    \label{2:3}     \\   
&&s\,(s\,\zeta')' 
=\frac{\kappa}{3}\,(-2A+2B-4C+D)
\, ,         \label{2:5} \\
&&s\,(s\,\mu')'
-{\rm e}^{2\nu+2\zeta+2\mu}
=\frac{\kappa}{3}\,(A-B-C-2D)
\, , \label{2:4} \\
&&s^2(\mu^{\prime 2}+2\nu'\mu'+2\zeta'\mu'+\zeta'\nu')
-{\rm e}^{2\nu+2\zeta+2\mu}  \nonumber \\
&&~~~~~~~~=\frac{\kappa}{2}\,(A+2B-2C-D) \, .  \label{2:6} 
\eea
Here
$A={\rm e}^{-2\zeta}s^2 H^{\prime 2}$,
$B={\rm e}^{-2\mu}s^2 w^{\prime 2}$,
$C={\rm e}^{2\nu+2\mu}w^2H^2$,
$D={\rm e}^{2\nu+2\zeta}(w^2-1)^2$ and 
$^\prime \equiv\frac{d}{dr}$.
In this system the last equation is the initial value ``Gauss'' constraint 
generating the residual gauge symmetry
$r\to\tilde{r}(r)$. This symmetry can be used to impose a 
gauge condition on the amplitudes.
The equations are also invariant with respect to the
rescalings of the $t$ and $x^4$ coordinates, 
$\nu\to \nu+\nu_0$,  
$\zeta\to\zeta+\zeta_0, H\to{\rm e}^{\zeta_0}H$, 
with constant $\nu_0$ and $\zeta_0$.

In addition, the equations are invariant with respect to the dilatations,  
$\mu\to\mu+\epsilon$, 
$\lambda\to\lambda+\epsilon$, 
$\zeta\to\zeta+\epsilon$, 
$\kappa\to{\rm e}^{2\epsilon}\kappa$, with constant $\epsilon$. 
The associated conserved  Noether charge  is
\be                 \label{noether}
{\cal Q}=s(-\nu'+\zeta'+\kappa HH'{\rm e}^{-2\zeta}).
\ee
The origin of this symmetry can be traced to the fact that 
effectively the system can be  viewed as 
the D=4 EYM-Higgs theory coupled to a dilaton. 
In general, 
if nothing depends on $x^4$, 
one can parameterize the 5-fields as $\g_{MN}dx^M dx^N=
{\rm e}^{-\zeta} \gamma_{\mu\nu}dx^\mu dx^\nu-
{\rm e}^{2\zeta}(dx^4)^2$ 
and 
$A^a_Mdx^M=A^a_\mu dx^\mu+\,H^adx^4$.   
The action (\ref{1}) then reduces to 
$S=\frac{1}{g^2}\int dx^4\int d^4x\sqrt{-\gamma}\,{\cal L}_4$ with 
\be                   \label{1+}
{\cal L}_4=-\frac{1}{2\kappa g^4}\,{\cal R}
+\frac{3}{\kappa g^4}\,\partial_\mu\zeta\partial^\mu\zeta
-\frac14\,{\rm e}^\zeta 
F^a_{\mu\nu}F^{a\mu\nu}
+\frac12\,{\rm e}^{-2\zeta} 
{\cal D}_{\mu}H^a {\cal D}^{\mu}H^a\, .
\ee
Here ${\cal R}$ is the Ricci scalar for  $\gamma_{\mu\nu}$,
the indices are lifted by  $\gamma^{\mu\nu}$, also 
$F^a_{\mu\nu}=\partial_\mu A^a_\nu-\partial_\nu A^a_\mu
+\varepsilon_{abc}A^b_\mu A^c_\nu$ 
and ${\cal D}_{\mu}H^a=\partial_\mu H^a+
\varepsilon_{abc}A^b_\mu H^c$. 
This determines the theory of coupled 
EYM-Higgs-dilaton fields in D=4.
The amplitude $\zeta$ therefore effectively plays the role of the 
dilaton,  which explains
the dilatational symmetry.
The vortex solutions under consideration thus can be viewed 
as D=4 gravitating YM-Higgs monopoles 
coupled to the dilaton in a special way  
(the systems considered so far in the literature  
do not have the direct coupling between the dilaton and Higgs fields
\cite{Brihaye01}.)

Some simplest solutions of the system (\ref{2:7})--(\ref{2:6}) can be found. 
The vacuum Schwarzschild solution is obtained for $w=\pm 1$, $H=0$, 
${\rm e}^{-2\lambda}=1-\frac{4\kappa M}{3r}$, and either $\nu=-\lambda$,
$\zeta=0$ or $\zeta=-\lambda$, $\nu=0$. 
The extreme charged Abelian  black string solution 
is described by
$w=0$, $H=H_0$, 
${\rm e}^{2\nu}={\rm e}^{2\zeta}={\rm e}^{-\lambda}=
1-r_H/r$,
with $r_H=\sqrt{2\kappa/3}$. 

We are interested in globally regular solutions with finite
mass per unit $x^4$. This
implies that $\nu=\lambda=\zeta=0$
at $r=\infty$, and if ${\rm e}^{\mu}=r$ then 
$\nu=-\frac{2\kappa}{3r}M+O(r^{-2})$ for large $r$,
while
$\lambda=\frac{\kappa }{3r}M_{\rm ADM}+O(r^{-2})$. 
Here $M$ and $M_{\rm ADM}$, respectively, are the Newtonian and 
the ADM masses, 
the dimensionful energy being $E=(4\pi/g^2)M$.
Choosing ${\rm e}^{\mu}=r$, dividing
Eq.(\ref{2:3}) by $(\kappa s)$ and integrating gives
\be               \label{MMM}
M=\int_0^\infty dr\,{\rm e}^{\nu}
\left(\frac{r^2}{2}\,{\rm e}^{-\zeta-\lambda}H^{\prime 2}\right. 
+\left.{\rm e}^{-\lambda}w^{\prime 2}
+{\rm e}^{\lambda-\zeta}w^2H^2
+{\rm e}^{\lambda+\zeta}\,
\frac{(w^2-1)^2}{2r^2}
\right).
\ee
In flat space, with $\nu=\lambda=\zeta=0$, the integrand here
can be rearranged as the sum of a total derivative
and two perfect squares. The latter vanish if the Bogomol'nyi 
equations hold,
$r^2H'+w^2-1=0$ and $w'+wH=0$. The BPS solution, 
$H=\coth r-1/r$ and $w=r/\sinh r$, describes the flat space vortex 
with $M=M_{\rm ADM}=1$.

The Noether charge ${\cal Q}$ vanishes for globally 
regular  solutions. As a result, using (\ref{noether}) 
with ${\cal Q}=0$ allows us to 
exclude $\nu$ from the equations. Next, introducing 
$h\equiv{\rm e}^{-\zeta}H$
the amplitude $\zeta$ can also be eliminated.  
Choosing the  radial gauge where $\lambda=0$ and defining 
$R\equiv{\rm e}^\mu$,  
the independent variables are $w$, $h$, $Z\equiv\zeta'$, and $R$. 
If these are determined, the amplitudes $\nu$ and $\zeta$ are given by
$\zeta=\int Zdr$, and 
$\nu=\zeta+\kappa\int (Zh^2+hh') dr$. 
The field equations (\ref{2:7})--(\ref{2:4})
can be reformulated as a seven-dimensional
autonomous system 
\be           \label{dyn}
\frac{d}{dr}\,y_k=F_k(y_s,\kappa),
\ee
with
$y_k=\{w,w',h,h',Z,R,R'\}$, where the functions $F_k(y_s,\kappa)$
can be read-off from  (\ref{2:7})--(\ref{2:4}). 
In addition, 
the constraint (\ref{2:6}) will restrict  
the initial values of the $y_k$'s. Eqs.(\ref{dyn}) possess
the dilatational symmetry
\be              \label{scale}
r\to\epsilon r, ~~~w\to w,~~~h\to h/\epsilon,~~~Z\to Z/\epsilon,~~~ 
R\to\epsilon R,~~~\kappa\to\epsilon^2\kappa,~~~ 
\ee
which also changes the mass as $M\to M/\epsilon $.
The following fixed points of the equations will
be important in our analysis:

I. { The origin,} $(w,h,Z,R)=(1,0,0,0)$. 
Here the constraint ``stable manifold'' of solutions that are regular 
for $r\to 0$  can be characterized by the Taylor expansions
\bea            \label{zero}
w&=&1-br^2+O(r^2),~~~ h=ar+O(r^3), \nonumber \\
Z&=&O(r^2),~~~ R=r+O(r^3).   
\eea

II. { Infinity,} $(w,h,Z,R)=(0,v,0,\infty)$, where the 
(unconstrained) stable manifold  can be
described by
\bea                        \label{inf}
&&w=A\,r^{Cv}{\rm e}^{-vr}+o({\rm e}^{-r}),
\ Z=\kappa{Q}{r^{-2}}+O(r^{-3}\ln r),\ \nonumber  \\
&&h=v(1-{C}{r^{-1}})+O(r^{-2}\ln r),  \\
&&R=r-m\ln r+m^2r^{-1}\ln r-r_\infty+{\gamma}{r^{-1}}
+O(r^{-2}\ln r)\, .  \nonumber
\eea
Here $m\equiv\kappa(Cv^2+(2+\kappa v^2)Q)$ determines the
ADM mass, $M_{\rm ADM}=\frac{3}{\kappa}\,m$, while  
$M=\frac12 M_{\rm ADM}-\frac32 Q$.
In (\ref{zero}),(\ref{inf}) 
$a$, $b$, $A$, $v$, $C$, $Q$, $r_\infty$, $\gamma$ 
are eight free parameters,
but due to the scaling symmetry (\ref{scale}) 
only seven are independent,
and we shall set $v=1$.  
\begin{figure}
\epsfxsize=9cm
\epsfysize=7cm
\centerline{\epsffile{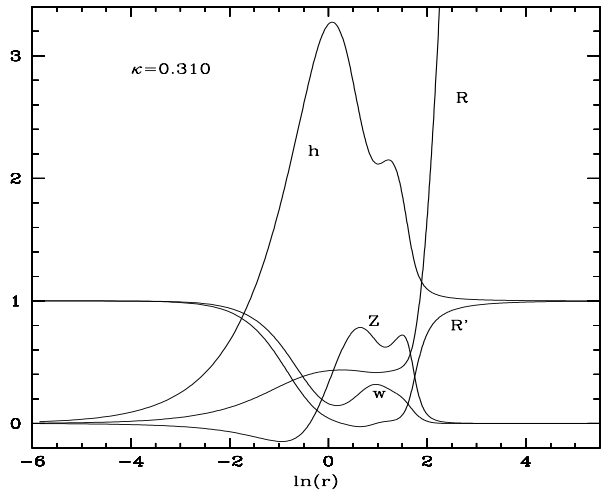}}
\caption{The vortex solution in the strong gravity regime.}
\label{fig1}
\end{figure}

III. {\sl  ``Warped'' $AdS_3\times S^2$.}
This fixed point is determined by the real root of
the algebraic equation
$4q^3+7q^2+11q=1$: 
\be                    \label{t}
w^2=q,~~
R^2=\kappa\frac{(11q-1)(1-q)}{(4q^2-13q+1)},~~
h^2=\frac{(1-q)}{R^2},~~ Z^2=-\frac{(4q^2-13q+1)}{(4q+1)R^2}. 
\ee
Evaluating, 
\be                \label{tube}
w=0.29,~~ h=\frac{1.27}{\sqrt{\kappa}},~~ 
Z=\pm\frac{0.31}{\sqrt{\kappa}} ,~~ 
R=0.75\sqrt{\kappa}\, .
\ee 
This is a new exact, essentially non-Abelian
solution to the field equations. Its geometry is 
\be               \label{ADS}
ds^2={\rm e}^{2\alpha Zr}dt^2-d r^2-{\rm e}^{2Zr}\,(dx^4)^2
-R^2\,d\Omega_2^2,
\ee 
with $\alpha=1+\kappa h^2=2.62$ 
(notice that for $\alpha=1$ this would be the metric
on $AdS_3\times S^2$). The gauge field is given by (\ref{gauge})
with $H={\rm e}^{Zr}h$. 
Linearizing Eqs.(\ref{dyn}) around this solution, 
one finds the 7 characteristic 
eigenvalues of the linearized system to be
$
\{-2.77,-2.47,-2.12,-0.61\pm 1.24\,i ,0.88,1.54\},
$
in units of $1/\sqrt{\kappa}$. 
The solution (\ref{t}) 
is therefore a hyperbolic (saddle) fixed 
point with five stable 
and two unstable for $r\to\infty$ eigenmodes.

We now integrate the equations (\ref{dyn})
for a given $\kappa\neq 0$ 
in the interval 
$r\in[0,\infty)$ using (\ref{zero}),(\ref{inf}) as the boundary
conditions and adjusting the seven independent free parameters to 
match the seven $y_k$'s. 
This gives the gravitating vortex solutions. For small $\kappa$ they are
only slightly deformed as compared to the flat space vortex.
The essentially non-Abelian field is contained in the
central core where $w\neq 0$, while in the far field 
zone $w$ vanishes and
only the long range U(1) component of the gauge field survives. 
The metric amplitudes are $Z\approx 0$, $R\approx r$,
$R'\approx 1$. 

It turns out to be convenient to measure the coupling to gravity 
not by $\kappa$ but by    
$\xi=a\kappa$ where $a$ comes from (\ref{zero}). 
Solutions are specified by their asymptotic parameters 
in (\ref{zero}),(\ref{inf}),  
which lie on the curve $\Gamma(\xi)$ 
in the eight-dimensional parameter
space with coordinates 
$\Gamma=(\kappa,a,b,A,C,Q,r_\infty,\gamma)$. 
As $\xi$ increases and $\Gamma(\xi)$ deviates from $\Gamma(0)$,
the solutions deviate more and more from their flat space profiles:
$R'(r)$ develops more and more pronounced 
minimum at some finite $r$ where $Z(r)$ considerably deviates
from zero; see Fig.\ref{fig1}. 
It is interesting that  $\kappa$ not always increases with $\xi$, 
but only up to the maximal value $\kappa_{\rm max}=3.22$ 
and then starts decreasing; 
see Fig.\ref{fig3} where projections of $\Gamma(\xi)$ on the $C\kappa$,
$CQ$, $CA$, and $C\gamma$ planes are shown.

For strongly gravitating solutions,  
as $\xi$ continues to increase, the amplitudes $w(r)$, 
$h(r)$, $Z(r)$, $R(r)$ start  developing
oscillations around the constant values which are
close to those given by Eq.(\ref{tube}). The number of 
oscillations grows as gravity gets stronger, 
while their amplitudes tend to zero. 
The proper length of the interval where oscillations take place 
increases, and solutions develop a long throat connecting the core of the 
vortex with the asymptotic region. 
In the strong gravity limit 
this throat gets infinitely long, and the vortex core 
becomes disconnected from the outside world. 
A similar phenomenon has been observed in D=4
\cite{Breitenlohner92,Breitenlohner95}. 

\begin{figure}
\epsfxsize=9cm
\epsfysize=7cm
\centerline{\epsffile{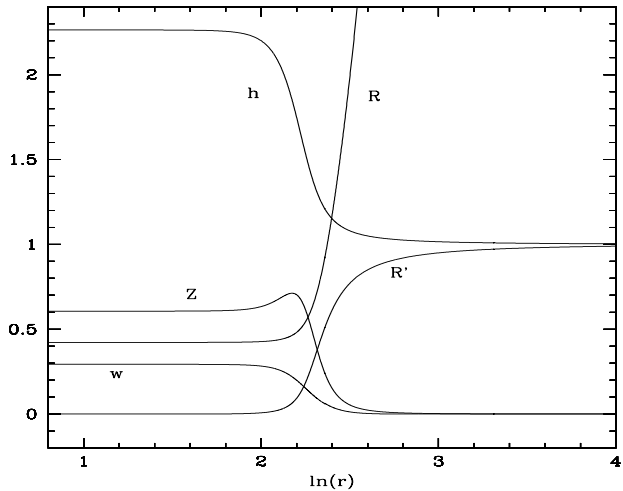}}
\caption{The exterior limiting solution. 
}
\label{fig2}
\end{figure}

In order to qualitatively understand such a behavior,  
each solution can be viewed as a trajectory interpolating in the 
phase space between the two fixed points (\ref{zero}) and (\ref{inf}). 
On its way, it gets attracted by the saddle point (\ref{t}), 
due to the five stable for $r\to\infty$ eigenmodes around this point,  
it spends some
``time'' in its vicinity, but finally it gets repelled 
due to the two unstable eigenmodes. As gravity gets stronger
with growing $\xi$, 
the trajectory approaches closer and closer 
the saddle point (\ref{t}) oscillating longer and longer 
in its vicinity. Finally,  the trajectory 
exactly hits the saddle point and splits into two parts. 
The first part starts at the origin (\ref{zero})
and after infinitely many oscillations 
arrives at the saddle point (\ref{t}). The second part
interpolates between the saddle point and infinity (\ref{inf}).
As a result, the vortex solution splits in the limit into two
independent solutions. The first, interior solution contains the regular core,
but asymptotically it is not flat and approaches instead the 
geometry (\ref{ADS}). The second, exterior 
solution interpolates between (\ref{ADS})
and the asymptotically flat region. 

\begin{figure}
\epsfxsize=9cm
\epsfysize=7cm
\centerline{\epsffile{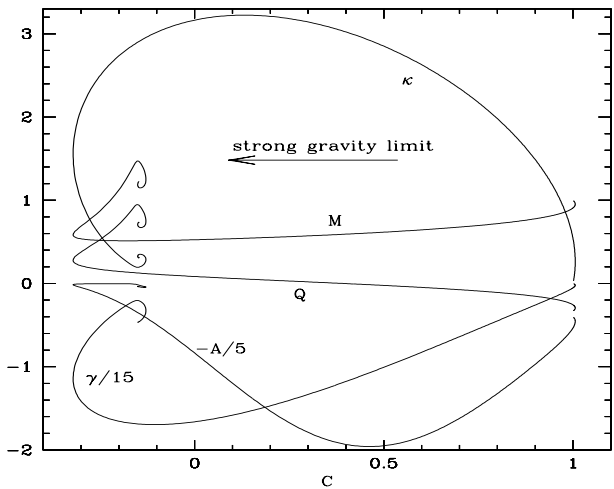}}
\caption{Parameters of the gravitating vortex solutions.
The right ends of the curves correspond to the flat space vortex, $\xi=0$. 
As $\xi$ increases, the curves extend to the left, and for
large $\xi$ they spiral towards the values corresponding to the limiting
solution.}
\label{fig3}
\end{figure}

One can 
directly construct the exterior limiting solution. 
This lives in the interval $r\in(-\infty,\infty)$.
Shifting $r$ to set $r_\infty=0$ leaves four
free coefficients in (\ref{inf}), while   
the (constraint) set of solutions that approach (\ref{tube}) for 
$r\to-\infty$ is determined by the two unstable modes 
around this fixed point. There are altogether six free parameters, 
and the matching conditions 
can be fulfilled if only $\kappa$ is treated
as the seventh free parameter. The solution is found for
$\kappa=0.316$; see Fig.\ref{fig2}.
 This solution can be viewed as an
 extreme 
non-Abelian black string: in $r=R$ coordinates 
${\rm e}^{-2\lambda}$ has a double zero at  
$r_{h}=0.42$, while 
${\rm e}^{2\nu}\sim (r-r_h)^{2.02}$ and  
${\rm e}^{2\zeta}\sim (r-r_h)^{0.77}$ for $r\to r_{h}$. 

It is instructive to consider the behavior of 
the parameters of the solutions 
$(\kappa,a,b,A,C,Q,r_\infty,\gamma)$ with increasing $\xi$. 
One can shift  $r\to r-r_\infty$ to set $r_\infty=0$, which changes
$A\to A{\rm e}^{r_\infty}$,
$\gamma\to \gamma+m r_\infty$ without affecting the other 
parameters. One can then plot all  parameters against $\xi$, but
it is  more illustrative to pick up one parameter, $C$, say,  
and plot the remaining ones against it. The resulting curves (some of them
are shown in Fig.\ref{fig3}) exhibit the characteristic spiraling 
behavior in the strong gravity limit (large $\xi$). 
A similar oscillatory behavior of parameters 
in the critical limit is known for relativistic/boson stars
\cite{star}.

Apart from the fundamental solutions, 
one finds also their excitations,  for which
$w$ oscillates around zero. These can be parameterized by the
number of nodes $n=1,2,\ldots $ of $w$.
These solutions 
do not have the flat space limit: for $\kappa\to 0$
their mass $M\sim1/\sqrt{\kappa}$. 
This limit can be analyzed with the rescaling (\ref{scale}) 
with $\epsilon=1/\sqrt{\kappa}$, which eliminates the explicit
$\kappa$-dependence from the equations. 
The rescaled Higgs field then decouples in the limit,  
and this leads to solutions which
resemble the D=4 Bartnik-McKinnon solutions
\cite{Bartnik88} -- with
$w$ oscillating around zero and $w(\infty)=\pm 1$. 
 
{\bf Acknowledgment--} I would like to thank 
Peter  Breitenlohner, Peter Forgacs, and especially Dieter Maison
for numerous clarifying discussions. This work was supported by the
CNRS and partly by the DFG grant Wi 777/4-3.


\begin{thebibliography}{1}
\bibitem{Aharony99}
O.~Aharony, S.~Gubser, J.~Maldacena, H.~Ooguri, and Y.~Oz,
\newblock {\em Phys.Rep.}, {\bbf 323},~183, 2000.
\newblock {\tt hep-th/9905111}.

\bibitem{Bartnik88}
R.~Bartnik, J.~McKinnon,
\newblock {\em Phys.Rev.Lett.}, {\bbf 61},~141, 1988.

\bibitem{Breitenlohner92}
P.~Breitenlohner, P.~Forgacs, D.~Maison,
\newblock {\em Nucl.Phys.}, {\bbf B 383},~357, 1992;
\newblock {\em Nucl.Phys.}, {\bbf B 442},~126, 1995;
K.~Lee, V.P.~Nair, E.J.~Weinberg, 
\newblock {\em Phys.Rev.}, {\bbf D 45},~2751, 1992.

\bibitem{Breitenlohner95}
P.~Breitenlohner, P.~Forgacs, D.~Maison,
\newblock {\em Comm. Math. Phys.}, {\bbf 163},~141, 1994; 
P.~Breitenlohner, D.~Maison,
\newblock {\em Comm.Math.Phys.}, {\bbf 171},~685, 1995.

\bibitem{Brihaye01}
Y.~Brihaye, B.~Hartmann, J.~Kunz, 
\newblock {\tt hep-th/0106227}. 



\bibitem{Gibbons94a}
G.W.~Gibbons, D.~Kastor, L.A.J.~London, P.K.~Townsend, J.~Traschen,
\newblock {\em Nucl.Phys.}, {\bf B 416},~850, 1994;
C.S.~Aulakh, D.~Maison, V.~Soni,
\newblock {\em Mod.Phys.Lett.}, {\bf A 9},~2139, 1994;
A.H. Chamseddine, M.S. Volkov,
\newblock {\em JHEP}, {\bf 0104},~023, 2001.

\bibitem{star}
B.K.~Harrison, K.S.~Thorne, M.~Wakano, J.A.~Wheeler,
\newblock {\em {\ssl Gravitation~Theory~and~Gravitational~Collapse}}.
\newblock Chicago, 1965; 
 R.~Friedberg, T.D.~Lee, Y.~Pang,
\newblock {\em Phys.Rev.}, {\bbf D 35},~3640, 1987.


\bibitem{Jaffe80}
A.~Jaffe,~C.H.~Taubes,
\newblock {\em {\ssl Vortices and Monopoles}}.
\newblock Birkh\"auser Boston, 1980.

\bibitem{Kastor92}
D.~Kastor, J.~Traschen,
\newblock {\em Phys.Rev.}, {\bbf D 46},~5399, 1992.

\bibitem{Randall99}
L.~Randall, R.~Sundrum,
\newblock {\em Phys.Rev.Lett.}, {\bbf 83},~4690, 1999.

\bibitem{Volkov98}
M.S. Volkov, D.V. Gal'tsov,
\newblock {\em Phys.Rep.}, {\bbf 319},~1, 1999;
M.S. Volkov, \newblock {\em Nucl.Phys.}, {\bbf B 566},~173, 2000.

\end{thebibliography}
\end{document}